\documentclass[prd,eqsecnum,twocolumn,amsfonts,showpacs]{revtex4}

\input epsf

\usepackage{graphicx}

\usepackage{bm}

\newcommand{\beq}{\begin{equation}}
\newcommand{\eeq}{\end{equation}}
\newcommand{\beqs}{\begin{eqnarray}}
\newcommand{\eeqs}{\end{eqnarray}}

\begin{document}

\title{Partition Function Zeros of a Restricted Potts Model on Lattice 
Strips and Effects of Boundary Conditions}

\author{Shu-Chiuan Chang$^{a}$ and Robert Shrock$^{b}$}

\bigskip

\affiliation{(a) \ Department of Physics \\
National Cheng Kung University \\
Tainan 70101, Taiwan} 

\bigskip

\affiliation{(b) \ C. N. Yang Institute for Theoretical Physics \\
State University of New York \\
Stony Brook, N. Y. 11794 }

\begin{abstract}

We calculate the partition function $Z(G,Q,v)$ of the $Q$-state Potts model
exactly for strips of the square and triangular lattices of various widths
$L_y$ and arbitrarily great lengths $L_x$, with a variety of boundary
conditions, and with $Q$ and $v$ restricted to satisfy conditions corresponding
to the ferromagnetic phase transition on the associated two-dimensional
lattices.  From these calculations, in the limit $L_x \to \infty$, we determine
the continuous accumulation loci ${\cal B}$ of the partition function zeros in
the $v$ and $Q$ planes.  Strips of the honeycomb lattice are also considered. 
We discuss some general features of these loci.


\end{abstract}

\maketitle

\newpage
\pagestyle{plain}
\pagenumbering{arabic}

\section{Introduction}

This is the second in a series of two papers on zeros of the $Q$-state Potts
model partition function \cite{wurev}-\cite{martinbook} on lattice strip graphs
of fixed width $L_y$ and arbitrarily great length $L_x$, with $Q$ and the
temperature-like variable $v$ restricted to satisfy the condition for the
ferromagnetic phase transition on the associated two-dimensional lattice. From
these calculations, in the limit $L_x \to \infty$, we exactly determine the
continuous accumulation loci ${\cal B}$ of the partition function zeros in the
$v$ and $Q$ planes.  These loci are determined by the equality in magnitude of
the eigenvalues of the transfer matrix of the model with maximal modulus and
hence are also called the set of equimodular curves (where ``curve'' is used in
a general sense that also includes line segments).  In the first paper
\cite{qvsdg} we carried out this study for self-dual strips of the square
lattice and found a systematic pattern of features, from which we were able to
conjecture properties applicable for strips of arbitrarily large widths.  Here
we continue this study by considering strips of the square, triangular, and
honeycomb lattices with a variety of different boundary conditions and studying
how these boundary conditions affect the loci ${\cal B}$.

We begin by briefly recalling the definition of the model and some relevant
notation. On a graph $G$ at temperature $T$, the Potts model is defined by the
partition function
\beq
Z(G,Q,v) = \sum_{ \{ \sigma \} }
\exp ( K\sum_{\langle i j \rangle} \delta_{\sigma_i \sigma_j} )
\label{z}
\eeq
where $\sigma_i=1,...,Q$ are the classical spin variables on each vertex (site)
$i \in G$, $\langle i j \rangle$ denotes pairs of adjacent vertices, $K=\beta
J$ where $\beta = (k_BT)^{-1}$, and $J$ is the spin-spin coupling.  We define
$a=e^K$ and $v=e^K-1$, so that $v$ has the physical range of values $0 \le v
\le \infty$ and $-1 \le v \le 0$ for the respective ferromagnetic and
antiferromagnetic cases $J > 0$ and $J < 0$.  The graph $G=G(V,E)$ is defined
by its vertex set $V$ and its edge (bond) set $E$. The number of vertices of
$G$ is denoted as $n=n(G)=|V|$ and the number of edges of $G$ as $e(G)=|E|$.
The Potts model can be generalized from non-negative integer $Q$ and physical
$v$ to real and, indeed, complex $Q$ and $v$ via the cluster relation \cite{kf}
$Z(G,Q,v) = \sum_{G^\prime \subseteq G}Q^{k(G^\prime)}v^{e(G^\prime)}$, where
$G'=(V,E^\prime)$ with $E^\prime \subseteq E$, and $k(G^\prime)$ denotes the
number of connected components of $G^\prime$.

The Potts model partition function is equivalent to an important function in
mathematical graph theory, the Tutte polynomial $T(G,x,y)$
\cite{wt1,wt2,biggsbook}:
\beq
Z(G,Q,v) = (x-1)^{k(G)}(y-1)^{n(G)} \, T(G,x,y)
\label{zt}
\eeq
where
\beq
x=1+\frac{Q}{v} \ , \quad y=1+v \ .
\label{xyqv}
\eeq

The phase transition temperatures of the ferromagnetic Potts model (in the
thermodynamic limit) on the square (sq), triangular (t), and honeycomb (hc) 
lattices are given, respectively, by the physical solutions to the equations 
\cite{wurev,roots} 
\beq
Q=v^2 \quad (sq) 
\label{qvsq}
\eeq
\beq
Q=v^2(v+3) \quad (t)
\label{qvtri}
\eeq
and
\beq
Q^2+3Qv-v^3=0 \quad (hc) 
\label{qvhc}
\eeq
The conditions (\ref{qvtri}) and (\ref{qvhc}) are equivalent, owing to the
duality of the triangular and honeycomb lattices.  In terms of the Tutte
variables, these conditions are $(y-1)(x-y)=0$ for $\Lambda=sq$,
$(y-1)(y^2+y-x-1)=0$ for $\Lambda=t$, and $(y-1)^2(x^2+x-y-1)=0$ for
$\Lambda=hc$.  Since $y=1$ means $K=0$, i.e., infinite temperature, the root at
$y=1$ is not relevant; dividing both sides of these three equations by the
appropriate powers of $(y-1)$, we thus obtain the conditions $x=y$ \quad (sq),
$y^2+y-x-1=0$ (t), and $x^2+x-y-1=0$ \cite{dual}.

Although the infinite-length, finite-width strips that we consider here are
quasi-one-dimensional systems and the free energy is analytic for all nonzero
temperatures, it is nevertheless of interest to investigate the properties of
the Potts model with the variables $Q$ and $v$ restricted to satisfy the above
conditions.  For a $L_y \times L_x$ section of the respective type of lattice,
as $L_x \to \infty$ and $L_y \to \infty$ with $L_y/L_x$ equal to a finite
nonzero constant, one sees the onset of two-dimensional critical behavior.  The
strips with $L_x \to \infty$ and $L_y$ fixed provide a type of interpolation
between the one-dimensional line and the usual two-dimensional thermodynamic
limit as $L_y$ increases. One of the interesting aspects of this interpolation,
and a motivation for our present study, is that one can obtain exact results
for the partition function $Z(G,Q,v)$ and (reduced) free energy $f=\lim_{n \to
\infty} n^{-1} \ln Z$.  The value of such exact results is clear since it has
not so far been possible to solve exactly for $f(\Lambda,Q,v)$ for arbitrary
$Q$ and $v$ on a lattice $\Lambda$ with dimensionality $d \ge 2$, and the only
exact solution for arbitrary $v$ is for the $d=2$ Ising case $Q=2$. Thus, exact
results on the model for infinite-length, finite-width strips complement the
standard set of approximate methods that are used for $d \ge 2$, such as series
expansions and Monte Carlo simulations. Although the singular locus ${\cal
B}_v$ does not intersect the real axis on the physical finite-temperature
interval $-1 < v < \infty$ for the infinite-length, finite-width strips under
consideration here, properties of the image of this locus under the above
mappings (\ref{qvsq})-(\ref{qvhc}), ${\cal B}_Q$, can give insight into the
corresponding locus ${\cal B}_Q$ for the respective two-dimensional lattices.
Indeed, one of the interesting results of the present work and our
Ref. \cite{qvsdg} on infinite-length, finite-width strips is the key role of
the point $Q=4$ for the locus ${\cal B}_Q$, which can make a connection with
the locus ${\cal B}_Q$ for the physical phase transition of the Potts model on
two-dimensional lattices.

We next describe the boundary conditions that we consider.  The longitudinal
and transverse directions of the lattice strip are taken to be horizontal (in
the $x$ direction) and vertical (in the $y$ direction). Boundary conditions
that are free, periodic, and periodic with reversed orientation are labelled
$F$, $P$, and $TP$.  We consider strips with the following types of boundary
conditions:
\begin{enumerate}

\item 

$(FBC_y,FBC_x)=$ free 

\item 

$(FBC_y, PBC_x)=$ cyclic (cyc.) 

\item 

$(FBC_y, TPBC_x)=$ M\"obius  (Mb.) 

\item 

$(PBC_y, FBC_x)=$ cylindrical  (cyl.) 

\item 

$(PBC_y, PBC_x)=$ toroidal (tor.) 

\item 

$(PBC_y, TPBC_x)=$ Klein-bottle (Kb.) 

\end{enumerate}
We thus denote a strip graph of a given type of lattice $\Lambda=sq$ or $t$
as $\Lambda[L_y \times L_x],BC$, where $BC=$free for $(FBC_y,FBC_x)$ and
similarly for the other boundary conditions.  In earlier work we showed that
although the partition functions $Z(\Lambda[L_y \times L_x],BC,Q,v)$ are
different for cyclic and M\"obius boundary conditions, ${\cal B}$ is the
same for these two, and separately that although this partition function is
different for toroidal and Klein-bottle boundary conditions, ${\cal B}$ is 
the same for these two latter conditions.  Therefore, we shall focus
here on the cases of free, cyclic, cylindrical, and toroidal boundary
conditions.  

Our procedure for calculating $Z(G,Q,v)$ on these strips is as follows.  For
the square and triangular lattices, the equations (\ref{qvsq}) and
(\ref{qvtri}) have the simplifying feature that they have the form
$Q=g_\Lambda(v)$, where $g_\Lambda(v)$ is a polynomial in $v$.  Accordingly, to
restrict $Q$ and $v$ to satisfy the phase transition conditions for the
respective two-dimensional lattices, we start with the exact partition function
and replace $Q$ by $g_\Lambda(v)$ for $\Lambda=sq$ and $\Lambda=t$.  We then
solve for the zeros of $Z(G,g_\Lambda(v),v)$ and, in the $L_x \to \infty$
limit, the continuous accumulation loci ${\cal B}_v$ in the complex $v$ plane.
The image of these zeros and loci under the respective mappings (\ref{qvsq})
and (\ref{qvtri}) yield the zeros and loci in the complex $Q$ plane. In the
case of the honeycomb (hc) lattice, the ferromagnetic phase transition
condition (\ref{qvhc}) is nonlinear in both $v$ and $Q$.  Since it is of lower
degree in $Q$, we solve for this variable, obtaining $Q=(v/2)( -3 \pm
\sqrt{9+4v} \, )$.  Of course, only one of these solutions is physical for the
actual two-dimensional lattice, namely the one with the plus sign.  Given the
fact that the triangular and honeycomb lattices are dual to each other and the
consequence that properties of the phase transition of the ferromagnetic Potts
model on the triangular lattice are simply related to those on the honeycomb
lattice, it follows that, insofar as we are interested in applying our exact
results on infinite-length, finite-width strips to two-dimensions, it suffices
to concentrate on strips of either the triangular or honeycomb lattice.  Since
the condition (\ref{qvtri}) is easier to implement than the solution for $Q$
given above, we shall mainly focus on strips of the triangular lattice, but
also include some comments on honeycomb-lattice strips.

We denote the Tutte-Beraha numbers \cite{wt1,wt2,bkw} 
\beq
Q_r = 4\cos^2(\pi/r) \ . 
\label{qr}
\eeq
For the range of interest here, $1 \le r \le \infty$, we note that $Q_r$
monotonically decreases from 4 to 0 as $r$ increases from 1 to 2, and then
$Q_r$ increases monotonically from 0 to 4 as $r$ increases from 2 to $\infty$. 
For our analysis of strips of the square lattice, it will also be useful to
denote $v_r = -2\cos(\pi/r)$ so that $Q_r = v_r^2$. Further background is given
in Ref. \cite{qvsdg}.

\section{Some General Structural Properties}

For these strips, the partition function has the general form (with $L_x=m$)  
\beq
Z(\Lambda[L_y \times m],BC,Q,v) = \sum_j c_j \, (\lambda_{\Lambda,BC,L_y,j})^m 
\label{zgsum}
\eeq
where the coefficients $c_j$ are independent of $m$.  It will be convenient to
separate out a power of $v$ and write
\beq
\lambda_{\Lambda,BC,L_y,j} = v^{L_y} \, \bar\lambda_{\Lambda,BC,L_y,j}
\label{lambdabar}
\eeq

We denote cyclic strips of the lattice $\Lambda$ of width $L_y$ and length $L_x
\equiv m$ as by $\Lambda[L_y \times m],cyc.$.  The partition function
has the general form \cite{saleur1,saleur2,cf}
\beq 
Z(\Lambda[L_y \times m],cyc.,Q,v) = \sum_{d=0}^{L_y} c^{(d)}
\sum_{j=1}^{n_Z(L_y,d)} (\lambda_{\Lambda,L_y,d,j})^m
\label{zgsumcyc}
\eeq
where we use simplified notation by setting $\lambda_{\Lambda,cyc.,L_y,d,j}
\equiv \lambda_{\Lambda,L_y,d,j}$ and $\bar\lambda_{\Lambda,cyc.,L_y,d,j}
\equiv \bar\lambda_{\Lambda,L_y,d,j}$, and where

\beq
n_Z(L_y,d) = \frac{(2d+1)}{(L_y+d+1)}{2L_y \choose L_y-d}
\label{nzlyd}
\eeq
for $0 \le d \le L_y$ and zero otherwise, and 
\beq
c^{(d)} = \sum_{j=0}^d (-1)^j {2d-j \choose j} Q^{d-j} \ . 
\label{cd}
\eeq
The coefficients $c^{(d)}$ can also be expressed in terms of Chebyshev
polynomials of the second kind as $c^{(d)}= U_{2d}(\frac{\sqrt{Q}}{2})$.  The
first few of these coefficients are $c^{(0)}=1$, $c^{(1)}=Q-1$,
$c^{(2)}=Q^2-3Q+1$, and $c^{(3)}=Q^3-5Q^2+6Q-1$.  The form (\ref{zgsumcyc})
applies for cyclic strips of not just the square lattice, but also the
triangular and honeycomb lattices \cite{cf,hca}.  The total number of
eigenvalues is
\beq
N_{Z,L_y,\lambda} = \sum_{d=0}^d n_Z(L_y,d) = {2L_y \choose L_y} \ . 
\label{nztot}
\eeq
Since $n_Z(L_y,L_y)=1$, i.e., there is only a single
$\lambda_{\Lambda,L_y,d,j}$ for $d=L_y$, we denote it simply as
$\lambda_{\Lambda,L_y,L_y}$. The single reduced eigenvalue with $d=L_y$ is
\beq
\bar\lambda_{\Lambda,L_y,L_y}=1  \ . 
\label{lamlast}
\eeq

  We now proceed with our results.  We shall point out relevant features for
the widths that we consider; of course, it is possible to study larger widths,
but, as our discussion will show, relevant features are already present for the
widths that we consider.

\section{Strips of the Square Lattice}

\subsection{Free} 

We denote these strips as $sq[L_y \times m],free$. For $L_y=1$, an elementary
calculation yields $Z=Q(Q+v)^{m-1}$.  Setting $Q=v^2$ yields
\beq
Z(sq[1 \times m],free,v^2,v) = v^{m+1}(v+1)^{m-1}
\label{zsqly1}
\eeq
which has zeros only at the two discrete points $v=0$ and $v=-1$.  In this case
the continuous ${\cal B}$ degenerates to these two points.

\begin{figure}[hbtp]
\centering
\leavevmode
\epsfxsize=2.4in
\begin{center}
\leavevmode
\epsffile{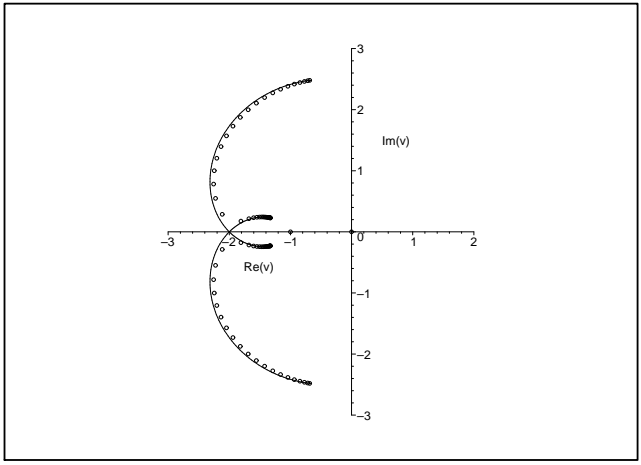}
\end{center}
\vspace{-10mm}
\caption{\footnotesize{Locus ${\cal B}_v$ for the Potts model on a $2 \times
\infty$ strip of the square lattice with free boundary conditions and with $Q$
and $v$ satisfying eq. (\ref{qvsq}).  Partition function zeros are shown for
a $2 \times 40$ strip.}}
\label{sqxy2v}
\end{figure}

\begin{figure}[hbtp]
\centering
\leavevmode
\epsfxsize=2.4in
\begin{center}
\leavevmode
\epsffile{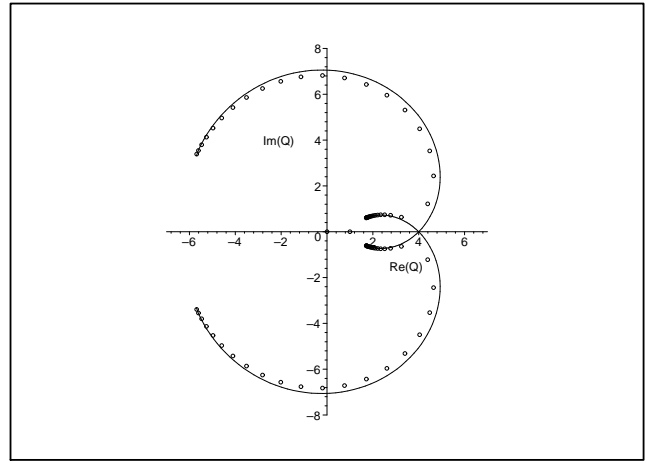}
\end{center}
\vspace{-10mm}
\caption{\footnotesize{Locus ${\cal B}_Q$ for the Potts model on a $2 \times
\infty$ strip of the square lattice with free boundary conditions and with $Q$
and $v$ satisfying eq. (\ref{qvsq}).  Partition function zeros are shown for
a $2 \times 40$ strip.}}
\label{sqxy2q}
\end{figure}

For $L_y=2$ we use the partition function $Z(sq[2 \times m],free,Q,v)$,
calculated in Ref. \cite{a}, which has the form (\ref{zgsum}) with two 
$\lambda$'s.  For $Q=v^2$, we find 
\beqs
\bar\lambda_{sq,2,0,j} & = & \frac{1}{2} \biggl [ (v+2)^2 \pm
(v^4+4v^3+12v^2+20v+12)^{1/2} \biggr ] \cr\cr
& & 
\label{lam2d0j12c}
\eeqs
where $j=1,2$ correspond to $\pm$.  In the infinite-length limit, the
continuous accumulation set loci ${\cal B}_v$ and, correspondingly, ${\cal
B}_Q$, are shown in Figs. \ref{sqxy2v} and \ref{sqxy2q}.  They consist of two
complex-conjugate arcs that intersect each other and cross the real $v$ axis at
$v=-2$ and equivalently, the real $Q$ axis at $Q=4$.  The endpoint of these
arcs in the $v$ plane occur at the roots of the polynomial in the square root
of $\bar\lambda_{sq,2,0,j}$, namely $v=-1.33 \pm 0.23i$ and $-0.67 \pm 2.48i$,
and hence $Q=1.71 \pm 0.61i$ and $Q=-5.71 \pm 3.34i$.  For comparison, in this
and other figures we show zeros of the partition function for long finite
strips; in this case, $m=40$. One sees that the zeros lie rather close to the
asymptotic loci ${\cal B}$ and that the density of zeros increases as one
approaches the endpoints of the arcs.  

In these figures and others shown below, there are also zeros of the partition
function that do not lie on the asymptotic accumulation loci. For example, in
general, for any graph $G$, the cluster relation given above shows that
$Z(G,Q,v)=0$ at the point $(Q,v)=(0,0)$, which lies on manifolds defined by all
of eqs. (\ref{qvsq})-(\ref{qvhc}).  Depending on the type of lattice strip
graph, this may or may not be an isolated zero or lie on the continuous
accumulation set of zeros, ${\cal B}$.  For example, for the $L_y=2$
square-lattice strips with free or cylindrical boundary conditions it is
isolated (cf. Figs. \ref{sqxy2v}-\ref{sqxpy2q}), while for the $L_y=2$ strips
with cyclic or toroidal boundary conditions, it lies on the loci ${\cal B}$ in
the $v$ and $Q$ planes (cf. Figs. \ref{sqpxy2v}-\ref{sqpxpy2q}) and similarly
for the triangular strips to be discussed below.  Another general result is
that for a graph $G$ with at least one edge, $Z(G,Q,v)=0$ at the point
$(Q,v)=(1,-1)$.  This follows because $Z(G,Q,-1)$, the partition function for
the zero-temperature Potts antiferromagnet, is precisely the chromatic
polynomial $P(G,Q)$, which counts the number of ways one can assign colors from
a set of $Q$ colors to the vertices of $G$, subject to the condition that no
two adjacent vertices have the same color. These are called proper colorings of
$G$.  Clearly, the number of these proper colorings of a graph $G$ vanishes if
it has at least one edge and there is only one color, i.e., if $Q=1$.  This
point $(Q,v)=(1,-1)$ is on the manifold defined by eq. (\ref{qvsq}) for the
square lattice (although not on the corresponding manifolds defined by
eqs. (\ref{qvtri}) and (\ref{qvhc}) for the triangular and honeycomb lattices).
Thus, one sees a zero at this point in the plots for the square-lattice
strips. In the cases we have studied, this zero is isolated.

For $L_y=3$ we use the exact calculation of $Z(sq[3 \times m],free,Q,v)$ for
general $Q$ and $v$ in Ref. \cite{s3a} and specialize to $Q=v^2$.  The
partition function depends on ($m$'th powers of) four eigenvalues which are
roots of a quartic equation (eq. (A.8) in Ref. \cite{s3a}).  In the limit
$L_x \to \infty$, ${\cal B}_v$ consists of complex-conjugate pairs of arcs, all
of which pass through the point $v=-2$, where all four roots of the above
quartic equation are degenerate in magnitude.  The arcs lying farthest from the
real axis cross the imaginary $v$ axis (at $v \simeq \pm 2.96i$).  Hence, the
imagine of this locus under the map $Q=v^2$ in the $Q$ plane, ${\cal B}_Q$ also
consist of complex-conjugate arcs which all pass through the point $Q=4$.
Furthermore, two of these cross the negative real $Q$ axis, so that ${\cal
B}_Q$ separates the $Q$ plane into regions.  There are also other zeros on the
negative real $v$ axis, and hence resultant zeros on the positive real $Q$
axis.  Using the calculations of $Z(sq[L_y \times m],free,Q,v)$ in
Ref. \cite{ts,zt}, we have performed the corresponding analyses for $L_y=4,5$
and have found similar features.

\subsection{Cylindrical} 

\begin{figure}[hbtp]
\centering
\leavevmode
\epsfxsize=2.4in
\begin{center}
\leavevmode
\epsffile{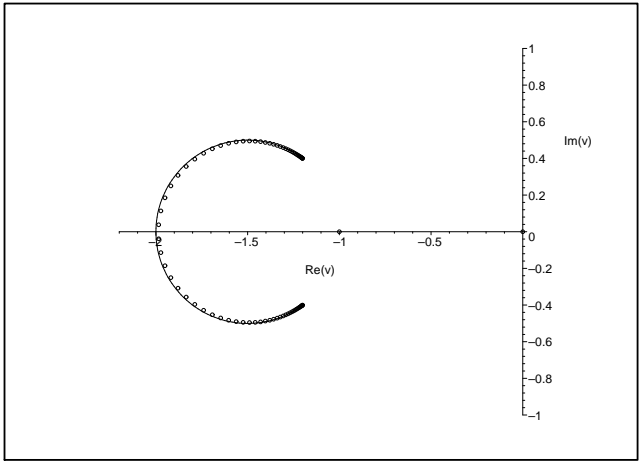}
\end{center}
\vspace{-10mm}
\caption{\footnotesize{Locus ${\cal B}_v$ for the Potts model on a $2 \times
\infty$ strip of the square lattice with cylindrical boundary conditions and
with $Q$ and $v$ satisfying the 2D phase transition condition (\ref{qvsq}).
Partition function zeros are shown for a cylindrical $2 \times 40$ strip.}}
\label{sqxpy2v}
\end{figure}

\begin{figure}[hbtp]
\centering
\leavevmode
\epsfxsize=2.4in
\begin{center}
\leavevmode
\epsffile{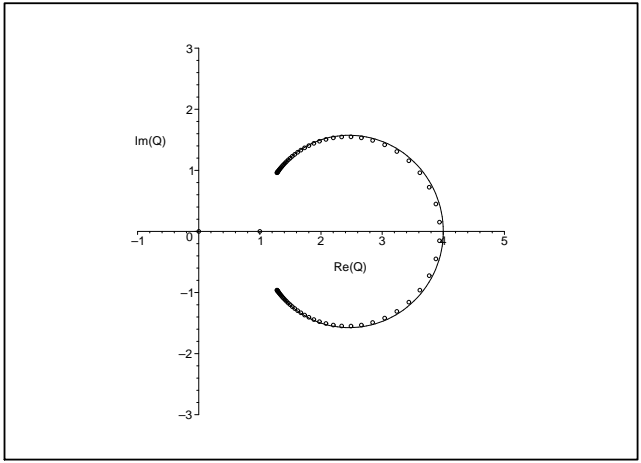}
\end{center}
\vspace{-10mm}
\caption{\footnotesize{Locus ${\cal B}_q$ for the Potts model on a $2 \times
\infty$ strip of the square lattice with cylindrical boundary conditions and
with $Q$ and $v$ satisfying eq. (\ref{qvsq}).  Partition function zeros are
shown for a cylindrical $2 \times 40$ strip.}}
\label{sqxpy2q}
\end{figure}

We find that $Z(sq[2 \times m],cyl.,v^2,v)$ has the form (\ref{zgsum})
depending on two $\bar\lambda$'s, which are 
\beqs
& & \bar\lambda_{sqcyl,j}=\frac{1}{2}\biggl [ 3v^2+8v+6 
\pm (v+2)\sqrt{5v^2+12v+8} \, \biggr ] \cr\cr
& & 
\label{lam_sqxpy2}
\eeqs
where $j=1,2$ correspond to $\pm$ and $sqcyl$ refers to this type of strip.  In
the limit $m \to \infty$, the locus ${\cal B}_v$ consists of a self-conjugate
arc that crosses the real axis at $v=-2$ and has endpoints at the roots of the
polynomial in the square root in eq. (\ref{lam_sqxpy2}), namely, $v=(-6 \pm
2i)/5$.  Thus, ${\cal B}_Q$ consists of an arc that crosses the real axis at
$Q=4$ and has endpoints at $Q=(32 \pm 24i)/25$.  These loci, together with
partition function zeros, are shown in Figs. \ref{sqxpy2v} and \ref{sqxpy2q}.
As was the case with the free strips, the density of zeros increases as one
approaches the endpoints of the arcs.  

\subsection{Cyclic and M\"obius}

For $L_y=1$, $sq[1 \times m]$ is just the circuit graph with $m$ vertices,
$C_m$.  An elementary calculation yields $Z(C_m,Q,v)=(Q+v)^m+c^{(1)}v^m$, so
for $Q=v^2$, one has $\bar\lambda_{sq,1,0}=v+1$ and $\bar\lambda_{sq,1,1}=1$ as
in eq. (\ref{lamlast}), and
\beq
Z(C_m,v^2,v)=v^m(v+1)[(v+1)^{m-1}+v-1] \ . 
\label{zsqpxy1}
\eeq
The resultant locus ${\cal B}_v$ is the circle $|v+1|=1$, i.e., 
\beq
v=-1+e^{i\phi} \ , \quad 0 \le \phi \le 2\pi
\label{vcirc}
\eeq
which crosses the real $v$ axis at $v=0$ and $v=-2$.  The resultant locus 
${\cal B}_Q$ with $Q=v^2$ is given by 
\beqs
Re(Q) & = & 2\cos \phi \ (\cos\phi-1) \ , \cr\cr
Im(Q) & = & 2\sin \phi \ (\cos\phi-1) 
\label{bqcirc}
\eeqs
for $0 \le \phi \le 2\pi$.  This locus crosses the real $Q$ axis at $Q=Q_2=0$,
where it has a cusp, and at $Q=Q_1=Q_\infty=4$; it also crosses the imaginary
$Q$ axis at $Q=\pm 2i$.  The loci ${\cal B}_v$ and ${\cal B}_Q$ divide the
respective $v$ and $Q$ planes each into two regions.  In the $v$ plane these
can be labelled as $R_1$ and $R_2$, the exterior and interior of the circle
$|v+1|=1$, and similarly in the $Q$ plane the exterior and interior of the
closed curve given by eq. (\ref{bqcirc}).  In regions $R_1$ and $R_2$ the
dominant $\bar\lambda$'s are $\bar\lambda_{sq,1,0}$ and $\bar\lambda_{sq,1,1}$,
respectively.

For $L_y=2$, we use the calculation of $Z(sq[2 \times m],cyc.,Q,v)$ in  
Ref. \cite{a}.  From
eq. (\ref{nzlyd}) we have $n_Z(2,0)=2$ and $n_Z(2,1)=3$, together with
$n_Z(2,2)=1$, for a total of $N_{Z,2,\lambda}=6$.  Specializing to the 
manifold of eq. (\ref{qvsq}), we find that 
\beq
\bar\lambda_{sq,2,1,1}=1+v
\label{lam2d1j1c}
\eeq
\beq
\bar\lambda_{sq,2,1,j}=v+2 \pm \sqrt{2v+3}
\label{lam2d1j23c}
\eeq
where $j=2,3$ correspond to $\pm$, and $\bar\lambda_{sq,2,0,j}$, given by
eq. (\ref{lam2d0j12c}).

\begin{figure}[hbtp]
\centering
\leavevmode
\epsfxsize=2.4in
\begin{center}
\leavevmode
\epsffile{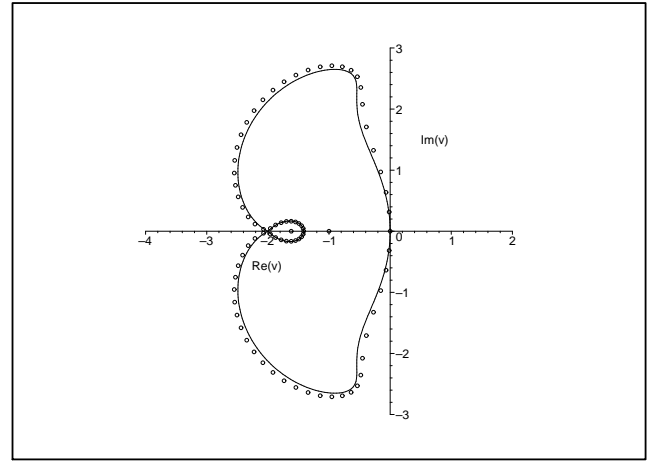}
\end{center}
\vspace{-10mm}
\caption{\footnotesize{Locus ${\cal B}_v$ for the Potts model on a $2 \times
\infty$ cyclic or M\"obius strip of the square lattice with $Q$ and $v$
satisfying eq. (\ref{qvsq}). Partition function zeros are shown for a cyclic 
$2 \times 40$ strip.}}
\label{sqpxy2v}
\end{figure}

\begin{figure}[hbtp]
\centering
\leavevmode
\epsfxsize=2.4in
\begin{center}
\leavevmode
\epsffile{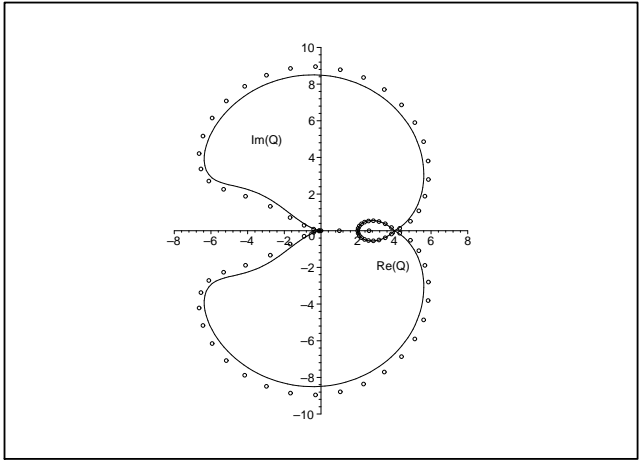}
\end{center}
\vspace{-10mm}
\caption{\footnotesize{Locus ${\cal B}_Q$ for the Potts model on a $2 \times
\infty$ cyclic or M\"obius strip of the square lattice with $Q$ and $v$
satisfying eq. (\ref{qvsq}). Partition function zeros are shown for a cyclic
$2 \times 40$ strip.}}
\label{sqpxy2q}
\end{figure}

For $m \to \infty$, the locus ${\cal B}_v$ for this cyclic (or corresponding
M\"obius) strip, shown in Fig. \ref{sqpxy2v}, is comprised of a single closed
curve that intersects the real $v$ axis at $v=0$ and $v=-\sqrt{2}$ and in a
two-fold multiple point at $v=-2$.  The image of ${\cal B}_v$ in the $q$ plane,
${\cal B}_Q$, shown in Fig.  \ref{sqpxy2q}, is again a closed curve that
crosses the real $Q$ axis once at $Q=0$ and $Q=2$ and in a two-fold multiple
point at $Q=4$, separating the complex $Q$ plane into three regions in 1-1
correspondence with those in the $v$ plane.  These regions are
\begin{itemize}

\item 

$R_1$, containing the real intervals $v > 0$ and $v < -2$ and extending outward
to the circle at infinity, and its image in the $Q$ plane, containing the real
intervals $Q \ge 4$ and $Q \le 0$, in which (with appropriate choice of the
branch cut for the square root in eq. (\ref{lam2d0j12c})) 
$\bar\lambda_{sq,2,0,1}$ is dominant, 

\item 

$R_2$, containing the real interval $-\sqrt{2} \le v \le 0$, and its image in
the $Q$ plane containing the real interval $0 \le Q \le 2$, in which 
$\bar\lambda_{sq,2,1,2}$ is dominant, 

\item 

$R_3$, containing the real interval $-2 \le v \le -\sqrt{2}$ and its image in
the $Q$ plane, containing the real interval $2 \le Q \le 4$, in which 
$\bar\lambda_{sq,2,2}=1$ is dominant. 

\end{itemize}
As was the case for the cyclic $L_y=1$ strip, the curve ${\cal B}_Q$ has a cusp
at $Q=0$.  For comparison with the asymptotic loci, in Figs. \ref{sqpxy2v} and
\ref{sqpxy2q} we also show partition function zeros calculated for a long
finite strip, with $m=40$. One sees that these lie close to the respective loci
${\cal B}$.

The exact calculation of $Z(sq[3 \times m],cyc.,Q,v)$ in Ref. \cite{s3a} has
the form of eq. (\ref{zgsumcyc}) with $L_y=3$.  From eq. (\ref{nzlyd}) we have
$n_Z(3,0)=5$ and $n_Z(3,1)=9$, $n_Z(3,2)=5$, together with $n_Z(3,3)=1$, for a
total of $N_{Z,3,\lambda}=20$.  We specialize to the manifold in
eq. (\ref{qvsq}).  Owing to the large number of $\bar\lambda_{sq,3,d,j}$'s, we
do not list them here.  In the $m \to \infty$ limit, we find that ${\cal B}_v$
crosses the real $v$ axis at $v=v_2=0$, $v=v_4=-\sqrt{2}$, $v=v_6=-\sqrt{3}$,
and $v=-2$, enclosing several regions in the $v$ plane.  The image locus under
the map (\ref{qvsq}), ${\cal B}_Q$, crosses the real axis in the interval $0
\le Q \le 4$ at $Q=Q_2=0$, $Q=Q_4=2$, $Q=Q_6=3$, and $Q=4$. It also crosses the
negative real axis at two points corresponding to the two pairs of
complex-conjugate points away from $v=0$ at which ${\cal B}_v$ crosses the
imaginary axis in the $v$ plane.  As with the $L_y=1$ and $L_y=2$ cyclic
strips, the curves on ${\cal B}_Q$ and ${\cal B}_v$ separate the respective $v$
and $Q$ planes in several regions in which different $\bar\lambda_{sq,3,d,j}$'s
are dominant.

We have performed corresponding calculations for the cyclic and M\"obius strips
of the square lattice with $L_y=4$ and $L_y=5$.  For brevity, we only comment
on ${\cal B}_Q$ here.  We find that ${\cal B}_Q$ crosses the real axis in the
interval $0 \le Q \le 4$ at $Q=4$ and at $Q=Q_{2\ell}$ for integer $1 \le \ell
\le L_y$ and also on the negative real axis. The outermost complex-conjugate
curves on ${\cal B}_Q$ continue the trend observed for smaller widths, of
moving farther away from the origin.  For example, the outermost curves on
${\cal B}_Q$ cross the imaginary $Q$ axis at $Q=\pm 2i$ for $L_y=1$, $Q \simeq
\pm 8.5i$ for $L_y=2$, and at progressively larger values for larger $L_y$.
Similarly, this outermost curve crosses the negative real axis farther away
from the origin; the approximate crossing point for $L_y=3$ is at $Q \simeq
-10$, with larger negative values for $L_y=4,5$.

\subsection{Toroidal and Klein-bottle} 

The exact solution for the partition function on the $L_y=2$ strip with
toroidal boundary conditions which we obtained in Ref. \cite{s3a} has the form
of eq. (\ref{zgsum}) with six $\lambda$'s.  Setting $Q=v^2$ (and using the
abbreviation $sqtor$ to indicate the boundary conditions), we find
\beq
\bar\lambda_{sqtor,2,j} = \bar\lambda_{sqcyl,2,j}
\label{lam20jc_sqpxpy2}
\eeq
where the $\bar\lambda_{sqcyl,2,j}$ with $j=1,2$ were given above in eq. 
(\ref{lam_sqxpy2}), 
\beqs
& & \bar\lambda_{sqtor,2,j} = \frac{1}{2}\biggl [ (v+2)(v+3) \cr\cr
& \pm & \Bigl [ (v^2+3v+8+4 \sqrt{2} \, )(v^2+3v+8-4\sqrt{2} \,) \Bigr 
]^{1/2} \biggr ] \ , \cr\cr
& & 
\eeqs
where $j=3,4$ correspond to the $\pm$ signs, 
\beq
\bar\lambda_{sqtor,2,5} = v+1 \ , 
\eeq
and
\beq
\bar\lambda_{sqtor,2,6}=1 \ . 
\eeq

\begin{figure}[hbtp]
\centering
\leavevmode
\epsfxsize=2.4in
\begin{center}
\leavevmode
\epsffile{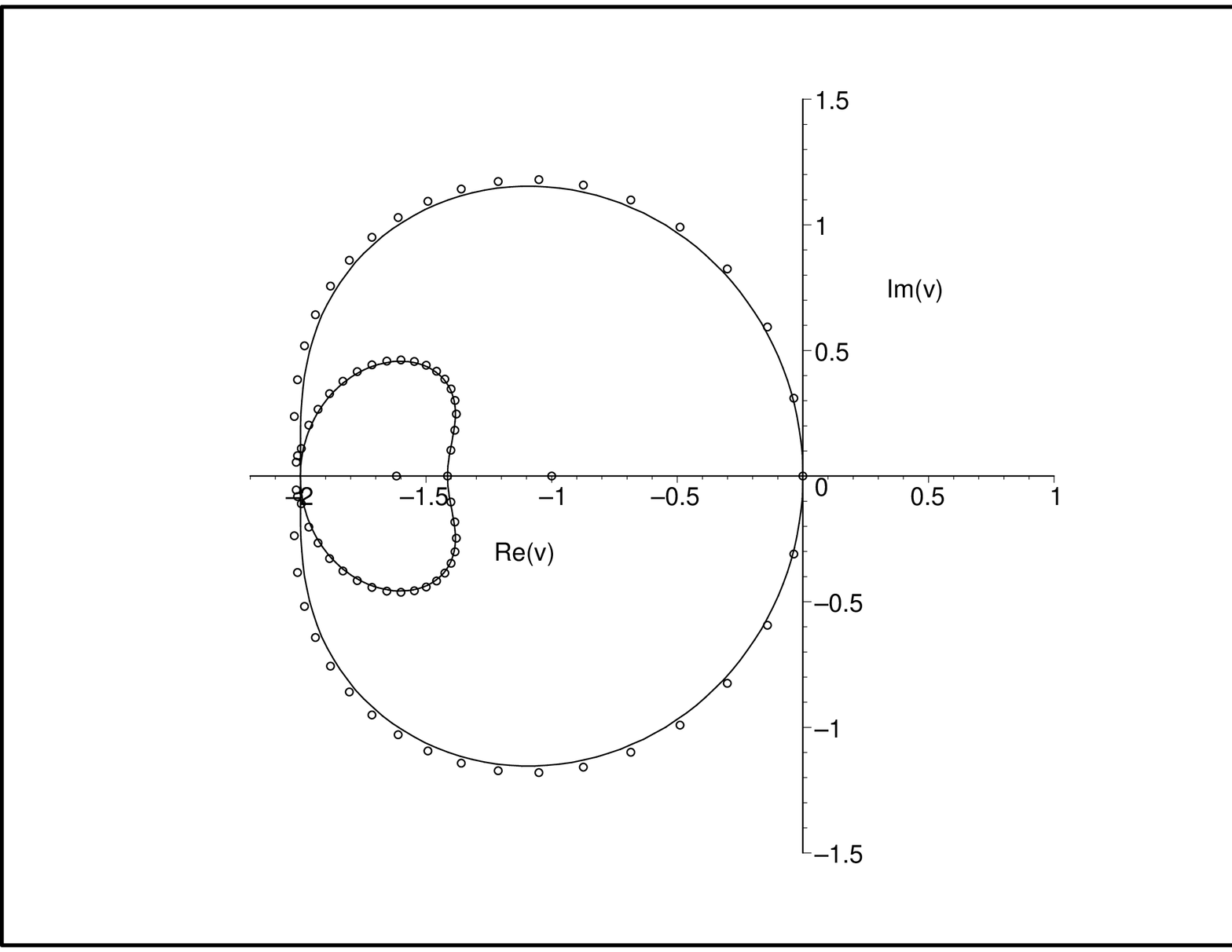}
\end{center}
\vspace{-10mm}
\caption{\footnotesize{Locus ${\cal B}_v$ for the Potts model on a $2 \times
\infty$ strip of the square lattice with toroidal or Klein-bottle boundary
conditions and with $Q$ and $v$ satisfying eq. (\ref{qvsq}).  Partition
function zeros are shown for a toroidal $2 \times 40$ strip.}}
\label{sqpxpy2v}
\end{figure}

\begin{figure}[hbtp]
\centering
\leavevmode
\epsfxsize=2.4in
\begin{center}
\leavevmode
\epsffile{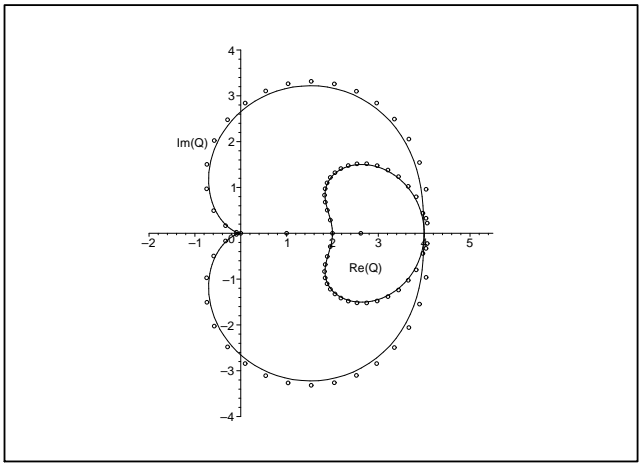}
\end{center}
\vspace{-10mm}
\caption{\footnotesize{Locus ${\cal B}_Q$ for the Potts model on a $2 \times
\infty$ strip of the square lattice with toroidal or Klein-bottle boundary
conditions and with $Q$ and $v$ satisfying (\ref{qvsq}).  Partition function
zeros are shown for a toroidal $2 \times 40$ strip.}}
\label{sqpxpy2q}
\end{figure}

In the $m \to \infty$ limit of this strip with toroidal or Klein-bottle
boundary conditions, we find that ${\cal B}_v$, shown in Fig. \ref{sqpxpy2v},
intersects the real $v$ axis at $v=0$ $v=-\sqrt{2}$, and $v=-2$. The image 
curve ${\cal B}_Q$, shown in Fig. \ref{sqpxpy2q}, thus intersects the real axis
at $Q=0,2,4$.  These curves divide the respective $v$ and $Q$ planes into three
regions.  In the $v$ plane, these are (i) the region $R_1$ including the
semi-infinite intervals $v \ge 0$ and $v \le -2$ and extending to complex
infinity, in which (with appropriate choices of branch cuts for the square
roots) $\bar\lambda_{sqtor,2,1}$ is dominant; (ii) the region $R_2$ including
the real interval $-\sqrt{2} \le v \le 0$, enclosed by the outer curve, in
which $\bar\lambda_{sqtor,2,3}$ is dominant, and (iii) the region $R_3$
enclosed by the innermost curve and including the real interval $-2 \le v \le
-\sqrt{2}$, in which $\bar\lambda_{sqtor,2,6}=1$ is dominant.  Corresponding
results hold in the $Q$ plane. 

Using our results in Ref. \cite{s3a,zttor}, we have also performed similar
calculations for the $L_y=3$ strip of the square lattice with toroidal boundary
conditions.  We find that ${\cal B}_v$ crosses the real axis at
$v=0$, $v=-\sqrt{2}$, $v=-2$, and $v \simeq -5.2$ and contains
complex-conjugate curves extending to complex infinity in the $Re(v) < 0$
half-plane.  Again, corresponding results hold in the $Q$ plane.

\section{Strips of the Triangular Lattice}

\subsection{General}

In order to investigate the lattice dependence of the loci ${\cal B}$, we have
also calculated these for infinite-length strips of the triangular lattice with
various boundary conditions and with $Q$ and $v$ restricted to satisfy the
phase transition condition for the two-dimensional triangular lattice,
eq. (\ref{qvtri}).  We construct a strip of the triangular lattice by starting
with a strip of the square lattice and adding edges connecting the vertices in,
say, the upper left to the lower right corners of each square to each other.
Since we will find that the values $Q_{2\ell}$ for $\ell=1,\cdots,L_y$ play an
important role for the loci ${\cal B}_Q$ for these strips with periodic
longitudinal boundary conditions, just as they did for the corresponding
square-lattice strips, we give a general solution of eq. (\ref{qvtri}) for the
case where $Q=Q_r$: the roots of this equation are $-1+2\cos(2(r+\eta)\pi)$
with $\eta = 1, \ -1, \ 0$, i.e., in order of increasing $v$,
\beq
v_{t1}(r) = -1 + 2\cos \biggl ( \frac{2(r+1)\pi}{3r} \biggr ) 
\label{vt1}
\eeq
\beq
v_{t2}(r) = -1 + 2\cos \biggl ( \frac{2(r-1)\pi}{3r} \biggr ) 
\label{vt2}
\eeq
and
\beq
v_{t3}(r) = -1 + 2\cos \biggl ( \frac{2\pi}{3r} \biggr )  \ . 
\label{vt3}
\eeq
The relevant case here is $r=2\ell$ with $1 \le \ell \le L_y$.  More generally,
for any real $r \ge 2$, these roots have the properties
\beq
v_{t1}(r) \le v_{t2}(r) \le 0 \quad {\rm for} \ \ r \ge 2 
\label{v12range}
\eeq
(where the first equality holds only at $r=\infty$ and the second equality
holds only at $r=2$), 
\beq
v_{t3}(r) \ge 0  \quad {\rm for} \ \ r \ge 2 
\label{v3range}
\eeq
(where the equality holds only at $r=2$).  As $r$ increases from 2 to $\infty$,
(i) $v_{t1}(r)$ increases from $-3$ to $-2$, (ii) $v_{t2}(r)$ decreases from 0
to $-2$, and (iii) $v_{t3}(r)$, which is the physical root for the phase
transition on the triangular lattice, increases from 0 to 1.  

For the cases of interest here, with $r=2\ell$ and $1 \le \ell \le L_y$, for
widths up to $L_y=5$, many of the trigonometric expressions in
eqs. (\ref{vt1})-(\ref{vt3}) simplify considerably, to algebraic expressions,
and in some cases to integers, so it is worthwhile displaying these roots
explicitly. For $r=2$ and $r=4$ we have
\beq
v_{t1}(2)=-3, \ v_{t2}(2)=0, \ v_{t3}(2)=0 \quad \Longrightarrow \quad Q=Q_2=0
\label{preimage_q0tri}
\eeq
\beqs
v_{t1}(4) & = & -1-\sqrt{3}, \ v_{t2}(4)=-1, \ v_{t3}(4)=-1+\sqrt{3} \cr\cr
& \Longrightarrow & \quad Q=Q_4=2
\label{preimage_q2tri}
\eeqs
We list the expressions for the $v_{tj}(r)$, $j=1,2,3$ for $r=6,8,10$ in the
Appendix.  For $r=\infty$, 
\beqs
& & v_{t1}(\infty)=v_{t2}(\infty)=-2, \ v_{t3}(\infty)=1 \cr\cr
& \Longrightarrow & \quad Q=Q_\infty=4
\label{vm2q4tri}
\eeqs
(Outside of our range, at $r=1$, since $Q_1=Q_\infty$, eq. (\ref{qvtri}) with
$Q=Q_1$ has the same set of roots as in (\ref{vm2q4tri}), with
$v_{t1}(1)=v_{t3}(1)=-2$ and $v_{t2}(1)=1$.)  Concerning the behavior of the
solutions of eq. (\ref{qvtri}) with $v$ as the independent variable, as $v$
decreases from 0, $Q$ increases from 0, reaching a maximum of 4 at $v=-2$ and
then decreasing through 0 to negative values as $v$ decreases through $-3$.
Thus, for all $v$ in the interval $-\infty \le v \le 0$, $Q$ is bounded above
by the value 4.

\subsection{Free}

The exact solution for the partition function on the $L_y=2$ strip of the
triangular lattice with free boundary conditions in Ref. \cite{ta} has the form
of eq. (\ref{zgsum}) with two $\lambda$'s.  Setting $Q=v^2(v+3)$ as in
eq. (\ref{qvtri}), we find the corresponding reduced $\bar\lambda$'s
\beqs
& & \bar\lambda_{t,2,0,j} = \frac{(v+1)}{2}\biggl [ v^3+5v^2+9v+7 
\cr\cr
& \pm & (v+3)\sqrt{(v+1)(v^3+3v^2+3v+5)} \ \biggr ]
\label{lam2tri_d0j12c}
\eeqs
where $j=1,2$ correspond to the $\pm$.

\begin{figure}[hbtp]
\centering
\leavevmode
\epsfxsize=2.4in
\begin{center}
\leavevmode
\epsffile{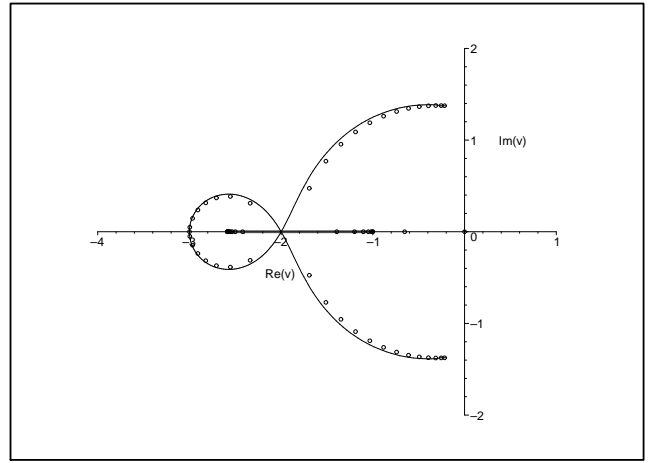}
\end{center}
\vspace{-10mm}
\caption{\footnotesize{Locus ${\cal B}_v$ for the Potts model on a $2 \times
\infty$ strip of the triangular lattice with free boundary 
conditions and with $Q$ and $v$ satisfying eq. (\ref{qvtri}). 
Partition function zeros are shown for a free $2 \times 20$ strip.}}
\label{txy2v}
\end{figure}

\begin{figure}[hbtp]
\centering
\leavevmode
\epsfxsize=2.4in
\begin{center}
\leavevmode
\epsffile{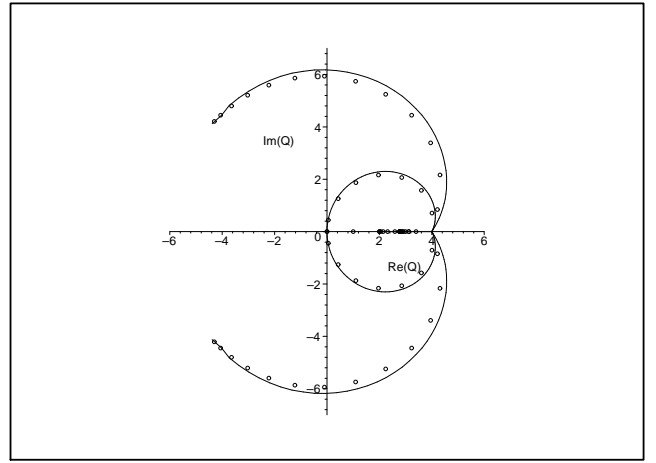}
\end{center}
\vspace{-10mm}
\caption{\footnotesize{Locus ${\cal B}_Q$ for the Potts model on a $2 \times
\infty$ strip of the triangular lattice with free boundary
conditions and with $Q$ and $v$ satisfying eq. (\ref{qvtri}). 
Partition function zeros are shown for a free $2 \times 20$ strip.}}
\label{txy2q}
\end{figure}

In the limit $m \to \infty$, we find the locus ${\cal B}_v$ shown in
Fig. \ref{txy2v} consisting of the union of (i) a curve that crosses the real
$v$ axis at $v=-2$ (a multiple point on the curve) and $v=-3$ and has endpoints
at the two complex-conjugate roots of the cubic factor in the square root in
eq. (\ref{lam2tri_d0j12c}), $v \simeq -0.206 \pm 1.37i$ and (ii) a line segment
on the real $v$ axis extending from $v=-1$ to $v \simeq -2.587$.  These
endpoints of the line segment are the other two zeros of the polynomial in the
square root in eq.  (\ref{lam2tri_d0j12c}).  This locus divides the complex $v$
plane into two regions.  The image of this locus under the mapping
(\ref{qvtri}), ${\cal B}_Q$, is shown in Fig. \ref{txy2q} and consists of the
union of a closed curve passing through $Q=0$ and $Q=4$ with complex-conjugate
arcs passing through $Q=4$ and terminating at endpoints $Q \simeq -4.38 \pm
4.12i$, and a line segment on the real axis extending from $Q=2$ to $Q \simeq
2.762$.  In Figs. \ref{txy2v} and \ref{txy2q} we also show zeros calculated for
a long finite free $L_y=2$ strip of the triangular lattice.  One sees again
that these lie close to the asymptotic loci ${\cal B}_v$ and ${\cal B}_Q$, as
one would expect for a long strip.  There are also discrete zeros that do not
lie on (or close to) the equimodular curves ${\cal B}$.  We have already
explained the origin of the zero at $(Q,v)=(0,0)$.  In contrast to the
situation with free strips of the square-lattice, where the zero at $v=0$ and
its image at $Q=0$ were both isolated, here we find that for the free $L_y=2$
strip of the triangular-lattice (see Figs. \ref{txy2v}, \ref{txy2q}), there is
an isolated zero at $v=0$, but because of the non one-to-one nature of the
mapping (\ref{qvtri}), its image in the $Q$ plane is not isolated but rather is
on ${\cal B}_Q$.  We have also performed analogous studies of wider strips of
the triangular lattice using the calculations of Ref. \cite{tt}, and we find
qualitatively similar results.

\subsection{Cyclic and M\"obius}
                   
For cyclic boundary conditions we have $N_{Z,2,\lambda}=6$,
$n_Z(2,0)=2$, $n_Z(2,1)=3$, $n_Z(2,2)=1$, just as in the square-lattice case.
We calculated the general partition function $Z(t[2 \times m],cyc.,Q,v)$ in
Ref. \cite{ta}, and this has the form of eq. (\ref{zgsumcyc}) with $L_y=2$. 
Restricting $Q$ and $v$ to satisfy eq. (\ref{qvtri}), we obtain
$\bar\lambda_{t,2,2}=1$, $\bar\lambda_{t,2,0,j}$ for $j=1,2$ given in 
eq. (\ref{lam2tri_d0j12c}), and $\bar\lambda_{t,2,1,j}$ for $j=1,2,3$, which
are solutions to the equation
\beqs
& & \eta^3 - (v+2)(3v+4)\eta^2 + (v+2)(v+1)(3v^2+9v+4)\eta \cr\cr
& &  -(v+1)^2(v^2+3v+1)^2=0 \ . 
\label{tri_cubeq_qv}
\eeqs

\begin{figure}[hbtp]
\centering
\leavevmode
\epsfxsize=2.4in
\begin{center}
\leavevmode
\epsffile{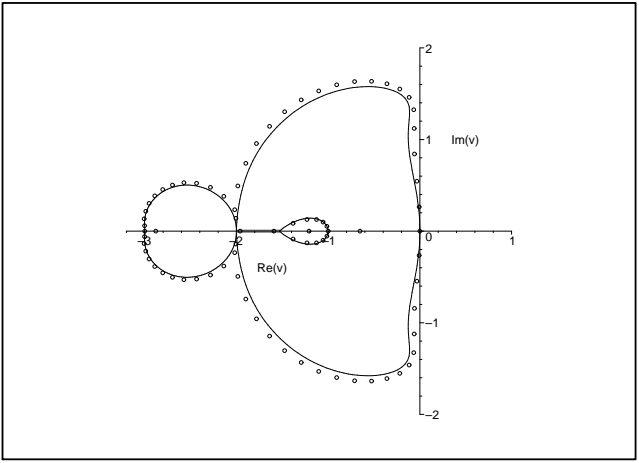}
\end{center}
\vspace{-10mm}
\caption{\footnotesize{Locus ${\cal B}_v$ for the Potts model on a $2 \times
\infty$ strip of the triangular lattice with cyclic or M\"obius boundary
conditions and with $Q$ and $v$ satisfying eq. (\ref{qvtri}). 
Partition function zeros are shown for a cyclic $2 \times 20$ strip.}}
\label{tpxy2v}
\end{figure}

\begin{figure}[hbtp]
\centering
\leavevmode
\epsfxsize=2.4in
\begin{center}
\leavevmode
\epsffile{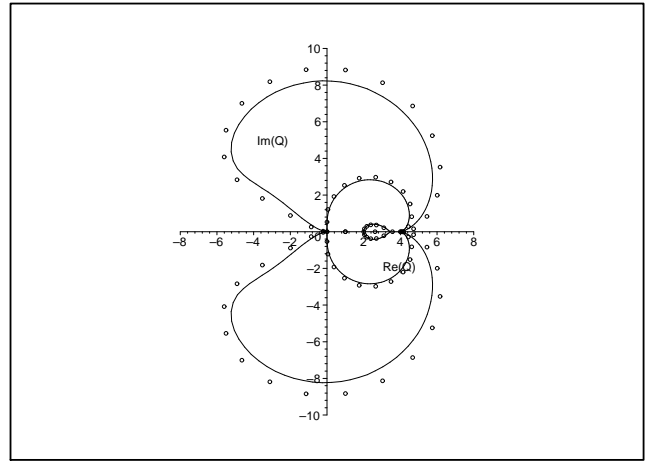}
\end{center}
\vspace{-10mm}
\caption{\footnotesize{Locus ${\cal B}_Q$ for the Potts model on a $2 \times
\infty$ strip of the triangular lattice with cyclic or M\"obius boundary
conditions and with $Q$ and $v$ satisfying eq. (\ref{qvtri}). 
Partition function zeros are shown for a cyclic $2 \times 20$ strip.}}
\label{tpxy2q}
\end{figure}

In the limit $m \to \infty$ for this cyclic strip (and for the same strip with
M\"obius boundary conditions), we find the locus ${\cal B}_v$ shown in
Fig. \ref{tpxy2v} and the corresponding locus ${\cal B}_Q$ shown in
Fig. \ref{tpxy2q}.  These loci consists of closed curves that intersect the
real $v$ and $Q$ axis at the following points, where the value of $Q$ is the
image of the value of $v$ under the mapping (\ref{qvtri}): (i) $v=0$ and
$v=-3, \ \Rightarrow \ Q=Q_2=0$, (ii) $v=-1, \ \Rightarrow \ Q=Q_4=2$, and
(iii) $v=-2, \ \Rightarrow \ Q=4$.  The locus ${\cal B}_v$ also contains a line
segment extending from $v \simeq -1.531$ to $v=-2$, and ${\cal B}_Q$ contains
its image under the mapping (\ref{qvtri}), extending from $Q \simeq 3.44$ to
$Q=4$.

The locus ${\cal B}_v$ separates the $v$ plane into four regions, as is evident
in Fig. \ref{tpxy2v}: 
\begin{itemize}

\item 

the region including the intervals $v \ge 0$ and $v \le -3$ on the real
$v$ axis and extending to complex infinity, in which (with appropriate
definition of the branch cuts associated with the square root in
eq. (\ref{lam2tri_d0j12c})) $\bar\lambda_{t,2,0,1}$ is dominant, 

\item 

the region including the neighborhood to the left of $v=0$ and excluding
the interior of the loop centered approximately around $v=-1.2$, in which one
of the $\bar\lambda_{t,2,1,j}$'s is dominant, 

\item 

the region in the interior of the loop centered around $v=-1.2$, in
which $\bar\lambda_{t,2,2}=1$ is dominant, 

\item 

the region including the interval that extends from $v=-2$ to $v=-3$, in which
another $\bar\lambda_{t,2,1,j}$ is dominant.

\end{itemize}

We give some specific dominant eigenvalues at special points.  At $v=0$,
$\bar\lambda_{t,2,0,1}=(7+3\sqrt{5})/2$, equal in magnitude to the dominant
$\bar\lambda_{t,2,1,j}$.  At $v=-2$, all of the six $\bar\lambda_{t,2,d,j}$'s
are equal in magnitude, and equal to unity. At $v=-3$,
$|\bar\lambda_{t,2,0,j}|=2$ for $j=1,2$, and two of the
$|\bar\lambda_{t,2,1,j}|=2$, while the third is equal in magnitude to 1.  

Correspondingly, the locus ${\cal B}_Q$ divides the $Q$ plane into five
regions:
\begin{itemize}

\item 

the region including the intervals $Q \ge 4$ and $Q \le 0$ on the real $Q$ 
axis and extending to infinity, in which $\bar\lambda_{t,2,0,1}$ is dominant 

\item 

two complex-conjugate regions bounded at large $Q$ by curves that cross the
imaginary axis at $Q \simeq \pm 8.23i$, in which one of the
$\bar\lambda_{t,2,1,j}$'s is dominant

\item 

the region including the interval $0 \le Q \le 2$, in which
another $\bar\lambda_{t,2,1,j}$ is dominant

\item 

the region in the interior of the loop centered approximately around $Q=2.6$,
in which $\bar\lambda_{t,2,2}=1$ is dominant

\end{itemize}

We have performed similar calculations for cyclic strips of the triangular
lattice with greater widths, $L_y=3,4,5$. We find that ${\cal B}_v$ crosses the
negative real axis at $v_{t2}(2\ell)$ for $1 \le \ell \le L_y$, so that ${\cal
B}_Q$ crosses the real axis at the image points under eq. (\ref{qvtri}),
$Q_{2\ell}$.  As with the cyclic square-lattice strips with widths $L_y \ge 3$,
we find that the outermost curves on ${\cal B}_Q$ cross the negative real axis;
for example, for $L_y=3$, such a crossing occurs at $Q \simeq -9.4$.

\subsection{Other Strips of the Triangular Lattice}

We have also calculated the partition function and the resultant loci ${\cal
B}_v$ and ${\cal B}_Q$ for strips of the triangular lattice with cylindrical
and toroidal boundary conditions.  A general feature that we find is that
${\cal B}_v$ passes through $v=-2$, and hence ${\cal B}_Q$ passes through
$Q=4$. Other features depend on the specific boundary conditions and width.
One property that we encounter is noncompactness of ${\cal B}_v$ and ${\cal
B}_Q$ (as was the case with the $L_y=3$ toroidal strip of the square lattice
and for cyclic self-dual strips of the square lattice \cite{qvsdg}).

\section{Strips of the Honeycomb Lattice} 

We first give a general solution for the three roots in $v$ of eq. 
(\ref{qvhc}) with $Q=Q_r$; in order of increasing value, these are 
\beq
v_{hc1}(r) = -4\cos \Bigl ( \frac{\pi}{r} \Bigr )
          \cos \biggl [ \frac{\pi}{3}\Bigl ( \frac{1}{r}-1 \Bigr )\biggr ]
\label{vhc1}
\eeq
\beq
v_{hc2}(r) = -4\cos \Bigl ( \frac{\pi}{r} \Bigr )
          \cos \biggl [ \frac{\pi}{3}\Bigl ( \frac{1}{r}+1 \Bigr )\biggr ]
\label{vhc2}
\eeq
and
\beq
v_{hc3}(r) = 4\cos \Bigl ( \frac{\pi}{r}  \Bigr ) 
              \cos \Bigl ( \frac{\pi}{3r} \Bigr ) \ . 
\label{vhc3}
\eeq
As $r$ increases from 2 to $\infty$, and $Q_r$ thus increases from 0 to 4, (i)
$v_{hc1}(r)$ decreases from 0 to a minimum of $-9/4$ at $r =
\pi/\arcsin(\sqrt{10}/8) \simeq 7.73$ and then increases to $-2$; (ii)
$v_{hc2}(r)$ decreases monotonically from 0 to $-2$; and (iii) $v_{hc3}(r)$,
the physical root, increases monotonically from 0 to 4. As with the analogous
expressions for the triangular lattice, it is straightforward to work out
simpler expressions to which eqs. (\ref{vhc1})-(\ref{vhc3}) reduce for special
values of $r$; we omit the details here.

For this honeycomb lattice, substituting the value $v=-2$ into
eq. (\ref{qvhc}) yields two solutions, $Q=2$ and $Q=4$.  Using our
general solutions for the partition functions $Z(hc[2 \times
m],free,Q,v)$ and $Z(hc[2 \times m],cyc.,Q,v)$ in Ref. \cite{hca}, we
have checked that at $v=-2$, $Q=4$ there is degeneracy of dominant
$\lambda$'s, so that these points are on the respective loci ${\cal
B}_v$ and ${\cal B}_Q$. This property is thus similar to the feature
that we have found for strips of the square and triangular lattices.
For $L_y=2$ cyclic honeycomb-lattice strips we find that, in addition,
the points $v=0$, $Q=Q_2=0$ and $v=-2$, $Q=Q_4=2$ are on the
respective loci ${\cal B}_v$ and ${\cal B}_Q$.  For the $L_y=3$ cyclic
strips of this lattice, ${\cal B}_v$ contains these points and also
$v_{hc1}(6)=-2\sqrt{3} \, \cos(5\pi/18)$, so that the image ${\cal
B}_Q$ contains the point $Q=Q_6=3$.  For all the strips of this
lattice that we have studied, with $L_y$ up to $L_y=5$, we find that
${\cal B}_v$ crosses the real axis at $v=v_{hc1}(2\ell)$ for
$\ell=1,\cdots,L_y$, so that ${\cal B}_Q$ crosses the real $Q$ axis at
$Q_{2\ell}$ for $\ell=1,\cdots,L_y$.

\section{Discussion}

We remark on several features that are common to all of the strips of the three
types of lattices that we have analyzed.  These include the following:
\begin{itemize}

\item 

For all of the strips, including those of the square, triangular, and
honeycomb lattices, that we have studied where nontrivial continuous
accumulation loci are defined (thus excluding the $1 \times m$ line
graph with free boundary conditions), we find that ${\cal B}_v$ passes
through $v=-2$ and ${\cal B}_Q$ passes through $Q=4$, which is the
image of $v=-2$ under both of the mappings (\ref{qvsq}) and
(\ref{qvtri}) and is a solution of eq. (\ref{qvhc}) with $v=-2$.

\item 

For all of the cyclic (and equivalently, M\"obius) strips that we have studied,
besides the crossing at $Q=4$, ${\cal B}_Q$ crosses the real axis at
\beq
Q=Q_{2\ell} \quad {\rm for} \ \ \ell=1,\cdots,L_y \ .
\label{q2ell}
\eeq
We conjecture that this holds for arbitrarily large $L_y$.  This locus can also
cross the real axis at other points, such as the crossings on the negative real
axis that we found for widths $L_y\ge 3$.

\end{itemize}
For the particular case of the cyclic square-lattice strips, these properties
agree with a result in Ref. \cite{saleur2}, namely, that at $Q=Q_{2\ell}$,
$|\lambda_{sq,L_y,\ell,max}|=|\lambda_{sq,L_y,\ell-1,max}|$, where
$\lambda_{sq,L_y,\ell,max}$ denotes the eigenvalue $\lambda_{sq,L_y,\ell,j}$ of
largest magnitude.  

We also observed that the points $v=-2$ and $Q=4$ play a special role for
self-dual strips of the square lattice in Ref. \cite{qvsdg}.  It is interesting
that for self-dual cyclic square-lattice strips, in addition to the point
$Q=4$, ${\cal B}_Q$ crosses the real axis at $Q_{2\ell+1}$ for $1 \le \ell \le
L_y$.  This set of points is interleaved with those in eq. (\ref{q2ell}).  As
we have noted in Ref. \cite{qvsdg}, these findings are in accord with the fact
that the Potts model at the values $Q=Q_r$ has special properties, such as the
feature that the Temperley-Lieb algebra is reducible at these values
\cite{martinber,saleur1,saleur2,ct}.  In Ref. \cite{qvsdg} we compared our
exact results for ${\cal B}$ on infinite-length, finite-width cyclic self-dual
strips of the square lattice with calculations of partition function zeros for
finite $L \times L$ sections of the square lattice with $Q=v^2$ in
Ref. \cite{kc} with the same boundary conditions. For finite lattice sections,
there is, of course, no locus ${\cal B}_Q$ defined, and hence one is only able
to make a rough comparison of patterns of zeros.  In the calculation of zeros
for the above-mentioned $L \times L$ section of the square lattice with cyclic
self-dual boundary conditions, e.g., for $L=8$, a number of zeros in the $Q$
plane occur at or near to certain $Q_r$'s, and the zeros in the $v$ and $Q$
planes exhibit patterns suggesting the importance of the points $v=-2$ and
$Q=4$.  With our results in the present paper, we can extend this comparison.
We see that the importance of $v=-2$ and $Q=4$ for the pattern of partition
function zeros for $Q$ and $v$ satisfying the relation (\ref{qvsq}) generalizes
to square-lattice strips with a variety of boundary conditions, not necessary
self-dual. Indeed, going further, our results show that the features we have
observed are true not just of the partition function on the square-lattice
strips with $Q=v^2$, but also on strips of the triangular and honeycomb
lattices with $Q$ and $v$ satisfying the analogous phase transition conditions
(\ref{qvtri}) and (\ref{qvhc}).  One interesting aspect of the findings in the
present work and our Ref. \cite{qvsdg} on infinite-length, finite-width strips
is the special role of the value $Q=4$ for the loci ${\cal B}_Q$, which can
make a connection with the locus ${\cal B}$ for the physical phase transition
of the Potts model on two-dimensional lattices.  In this context, we recall
that the value $Q=4$ is the upper end of the interval $0 \le Q \le 4$ for which
the ferromagnetic Potts model has a second-order transition on two-dimensional
lattices.

\section{Conclusions}

In conclusion, we have presented exact results for the continuous accumulation
set ${\cal B}$ of the locus of zeros of the Potts model partition function for
the infinite-length limits of strips of the square, triangular, and honeycomb
lattices with various widths, a variety of boundary conditions, and $Q$ and $v$
restricted to satisfy the conditions (\ref{qvsq}), (\ref{qvtri}), and
(\ref{qvhc}) for the ferromagnetic phase transition on the corresponding
two-dimensional lattices.  We have discussed some interesting general features
of these loci.

\section{Acknowledgments}

This research was partially supported by the Taiwan NSC grant
NSC-94-2112-M-006-013 (S.-C.C.) and the U.S. NSF grant PHY-00-98527 (R.S.).

\section{Appendix}

In this Appendix we list the $v_{tj}(r)$ for $r=6$, 8, and 10, which are
relevant to the analysis of the cyclic strips of the triangular lattice for
widths up to $L_y=5$.  From eqs. (\ref{vt1})-(\ref{vt3}), we have 
\beq
v_{tj}(6), \ j=1,2,3 \quad \Longrightarrow \quad Q=Q_6=3
\label{preimage_q3tri}
\eeq
where \cite{r6} 
\beq
v_{t1}(6) = -1+2\cos(7\pi/9) \simeq -2.532089
\label{vQ6c}
\eeq
\beq
v_{t2}(6) = -1+2\cos(5\pi/9) \simeq -1.347296
\label{vQ6b}
\eeq
\beq
v_{t3}(6) = -1+2\cos(\pi/9) \simeq 0.879385 
\label{vQ6a}
\eeq
\beq 
v_{tj}(8), \ j=1,2,3 \quad \Longrightarrow \quad Q=Q_8=2+\sqrt{2}
\label{preimage_Q8tri} 
\eeq
where
\beq
v_{t1}(8) = -1-\sqrt{2} \simeq -2.414214
\label{vQ8c}
\eeq
\beq
v_{t2}(8) = -1 + \frac{1-\sqrt{3}}{\sqrt{2}} \simeq -1.517638
\label{vQ8b}
\eeq
\beq
v_{t3}(8) = -1 + \frac{1+\sqrt{3}}{\sqrt{2}} \simeq 0.931852
\label{vQ8a}
\eeq
\beq 
v_{tj}(10), \ j=1,2,3 \quad \Longrightarrow \quad 
Q=Q_{10}=\frac{5+\sqrt{5} \,}{2}
\label{preimage_Q10tri} 
\eeq
where
\beqs
& & v_{t1}(10) = \frac{1}{4}\biggl (-5+\sqrt{5} \, - \sqrt{30+6\sqrt{5}} \ 
\biggr ) \simeq -2.338261 \cr\cr
& & 
\label{vQ10c}
\eeqs
\beqs
& & v_{t2}(10) = -\frac{(1+\sqrt{5} \, )}{2} \simeq -1.618034
\label{vQ10a}
\eeqs
\beqs
& & v_{t3}(10) = \frac{1}{4}\biggl (-5+\sqrt{5} \, + \sqrt{30+6\sqrt{5}} \ 
\biggr ) \simeq 0.956295 \cr\cr
& & 
\label{vQ10b}
\eeqs

\vfill
\eject


\begin{thebibliography}{99}
\bibitem{roots} 
We note that in the application of the conditions for the Potts
model on the two-dimensional lattices, the variable $Q$ is normally fixed (to a
non-negative integer) and one solves for the temperature-like variable $v$.
For each of the three lattices, although the equations
(\ref{qvsq})-(\ref{qvhc}) are nonlinear in $v$, only one of the (two, for
$\Lambda=sq$, three, for $\Lambda=t,hc$) roots in $v$ is in the physical
ferromagnetic range $0 \le v \le \infty$.
\bibitem{dual}
For a planar graph $G$, the Tutte polynomial satisfies $T(G,x,y)=T(G^*,y,x)$,
where $G^*$ is the planar dual of $G$.  The self-duality of the square lattice
is manifested in the fact that the condition $x=y$ is invariant under the
interchange $x \leftrightarrow y$.  The fact that the honeycomb lattice is the
dual of the triangular lattice is manifested in the property that the
conditions $y^2+y-x-1$ (t) and $x^2+x-y-1$ (hc) transform into each other under
the interchange $x \leftrightarrow y$.
\bibitem{ct} 
%
A related fact is that complex-temperature phase diagrams of the Potts model
show special properties at $Q=Q_r$. In addition to the trivial cases $Q=Q_2=0$
and $Q=Q_3=1$ and the exactly solvable Ising case $Q=Q_4=2$,
complex-temperature phase diagrams have been studied for various $Q_r$ such as
$Q_6=3$ and $Q_1=Q_\infty=4$ in, e.g., Refs. \cite{martinbook},
\cite{a}-\cite{tt} and \cite{chw}-\cite{jrs}.  These studies are complementary
to Refs. \cite{qvsdg,kc} and the present work in that $Q$ and $v$ are kept as
independent variables rather than being restricted to satisfy a condition such
as one of eqs. (\ref{qvsq})-(\ref{qvhc}).
\bibitem{r6} 
There are various equivalent ways to write the expressions for the roots of
eqs. (\ref{qvtri}) and (\ref{qvhc}) with $Q=Q_r$.  For example, for $r=6$,
i.e., $Q=Q_6=3$, the special cases of our general eqs. (\ref{vt1}),
(\ref{vt2}), and (\ref{vt3}) for $v_{tj}$, $j=1,2,3$, are equivalent to
eqs. (2.14), (2.13), and (2.12) of Ref. \cite{p2} and the special cases of our
general eqs. (\ref{vhc1}), (\ref{vhc2}), and (\ref{vhc3}) for $v_{hcj}$,
$j=1,2,3$, are equivalent to eqs. (3.5), (3.4), and (3.3) of Ref. \cite{p}.
\end{thebibliography}
\end{document}